# AKIBoards: A Structure-Following Multiagent System for Predicting Acute Kidney Injury

DAVID GORDON[1][0000-0002-5273-8344], PANAYIOTIS PETOUSIS[1][0000-0002-0696-608X], SUSANNE B. NICHOLAS [1][0000-0003-3535-9120], ALEX A.T. BUI[1][0000-0002-4702-1373]

[1] University of California, Los Angeles
`d.gordon@ucla.edu`

**Abstract.** Diagnostic reasoning entails a physician's local (mental) model based on an assumed or known shared perspective (global model) to explain patient observations with evidence assigned towards a clinical assessment. But in several (complex) medical situations, multiple experts work together as a team to optimize health evaluation and decision-making by leveraging different perspectives. Such consensus-driven reasoning reflects individual knowledge contributing toward a broader perspective on the patient. In this light, we introduce *STRUCture-following for Multiagent Systems (STRUC-MAS),* a framework automating the learning of these global models and their incorporation as prior beliefs for agents in multiagent systems (MAS) to follow. We demonstrate proof of concept with a prosocial MAS application for predicting acute kidney injuries (AKIs). In this case, we found that incorporating a global structure enabled multiple agents to achieve better performance (average precision, AP) in predicting AKI 48 hours before onset (structure-following-fine-tuned, SF-FT, AP=0.195; SF-FT-retrieval-augmented generation, SF-FT-RAG, AP=0.194) vs. baseline (non-structure-following-FT, NSF-FT, AP=0.141; NSF-FT-RAG, AP=0.180) for balanced precision-weighted-recall-weighted voting. Markedly, SF-FT agents with higher recall scores reported lower confidence levels in the initial round on true positive and false negative cases. But after explicit interactions, their confidence in their decisions increased (suggesting reinforced belief). In contrast, the SF-FT agent with the lowest recall decreased its confidence in true positive and false negative cases (suggesting a new belief). This approach suggests that learning and leveraging global structures in MAS is necessary prior to achieving competitive classification and diagnostic reasoning performance.

## 1 Introduction

When performing rounds in a hospital, physicians evaluate patients and communicate their clinical reasoning with their peers, who may agree or disagree with their assessment. Similarly, when physicians participate in clinical boards (e.g., tumor board for reviewing potential cancers and treatment), experts collaboratively form an assessment and assign a diagnosis, with the goal of exploring the full range of possibilities. Both types of clinical interactions involve multiple perspectives that are used to determine a patient's "next step." Such approaches extend to other interdisciplinary scenarios, where generalists (e.g., internists) often consult with others for specific expertise (e.g., nephrologists) [1]. These approaches aim to optimize health outcomes and diagnostic reasoning by employing different types of experience, coalescing individual insights into a more complete picture of the patient that informs clinical reasoning [2]. By analogy, local knowledge is woven together into an explanation for a patient based on a shared knowledgebase (a "global structure") representing all patients. These collective assessments have been shown to improve patient care and outcomes in myriad settings [3-5].

However, despite its high value, there can be significant costs and difficulties associated with this approach [6]: not all healthcare environments have the same degree of access to expertise (e.g., a limited number or no experts in a low-resource setting) [1, 4]; communication may be challenging (e.g., due to lack of (quality) documentation to understand another physician's reasoning); and decision-making may be time sensitive [7]. Indeed, the use of clinical boards is underutilized in part because of the need for considerable human resources.

To help address these issues, we introduce the first implementation of *STRUCture-following for Multiagent Systems* (STRUC-MAS) [8], a framework that draws inspiration from the construct of a clinical group of experts in a prosocial context [9] to facilitate "on-demand" specialist knowledge workers. Specifically, STRUC-MAS provides a way to learn the "global structure" of a domain problem, which can then be incorporated as prior beliefs for agents in multiagent systems (MAS) to follow. To demonstrate, we address the critical problem of predicting the onset of acute kidney injury (AKI) as a group of interacting agents, each with different perspectives (i.e., AKIBoards) that update their beliefs over time to reach agreement in an explainable manner [8].

## 2 Methods

### 2.1 AKIBoards: Predicting acute kidney injuries

Fig. 1 overviews the general architecture of STRUC-MAS [8], which was adapted to the specific problem of predicting acute kidney injuries. From a health perspective, AKIs can lead to chronic kidney disease (CKD) and other long-term health complications. Most AKIs are preventable if there is sufficient time to intervene and provide treatment. However, the onset of many AKIs go undetected until too late, and some are not even recognized until after the fact. To address this problem, AKIBoards was designed as a data-driven, clinically-oriented board using the STRUC-MAS framework. Following Stein et al.'s definition of a citizen-centric multiagent system, we shaped AKIBoards to be: 1) citizen-aware, taking into account stakeholder (health system, clinician) perspectives and requirements; 2) citizen-beneficial, providing value by improving current processes positively; and citizen-auditable, provide interpretable/explainable output and allow input from stakeholders [10].

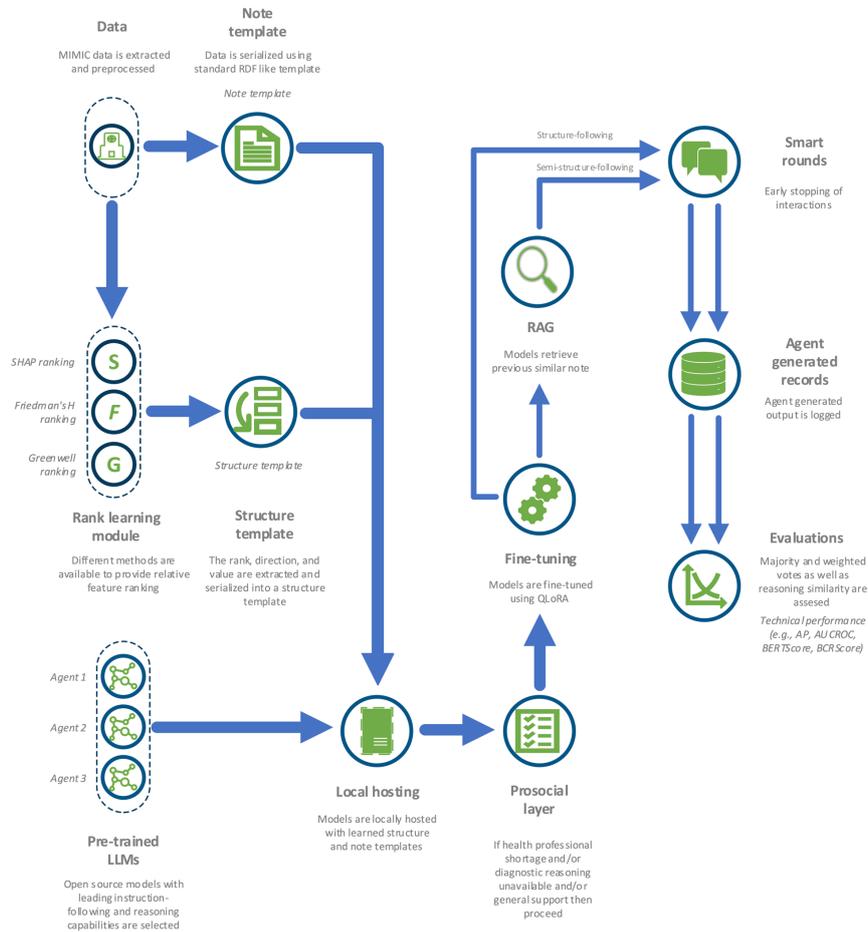

**Fig. 1.** AKIBoards in STRUC-MAS. Models are locally hosted with note and structure template. If the criteria from the prosocial layer are met, the framework proceeds with structure-following or semi-structure-following. Agent generated output is logged for evaluation.

**Dataset and representation.** To develop AKIBoards, an AKI dataset was extracted from MIMIC-III [11], a freely-available database comprising deidentified health-related data [11]. Details regarding the dataset as well as the developed AKI algorithm can be found in a previous study [12]. Briefly, the dataset contains clinical laboratory values and was stratified and randomized into train (70%, n=9,176), valid (15%, n=1,966), and test (15%, n=1,967) sets. The fine-tuning of a pre-trained large language model (LLM) was fit on the training set, the structures fit on the validation set, and the agents evaluated on the holdout test set. We adapted a basic note template, where the data is serialized into knowledge triples (*feature name*, *is*, *value*) [13]. Notably, serialization transforms the tabular data into a standardized note.

**Structure learning: Rank learning module and structure template.** When the structure is unknown to clinicians and/or agents, traditional machine learning or statistics methods can help learn plausible structures to follow [12]. We build on the rank learning module from Ranking Approaches for Unknown Structures (RAUS) [14] to learn structures that also capture feature value directionality and pairwise interactions [15-17]. In this work, we used SHapley Additive exPlanations (SHAP) given its wide acceptance as an interpretability method. We transformed categorical features into dummy variables to enable interpretability at the bin level. Further, we used best-k (i.e., top) ranked features, where k=10 (Appendix 6.1, Figure 2). Note that SHAP output is automatically incorporated into the standardized structure template (see Appendix 6.1, Table 1).

**Prosocial layer.** We integrate prosocial logic [9] via Boolean constraints to help ensure that the system is used for augmenting physicians rather than replacing them. The key issues this system aims to address are: 1) health professional shortages, either due to time constraints or unavailability; 2) unavailable reasoning for an AKI diagnosis, or to further support clinical decision making; and/or 3) general support (e.g., training of physicians to detect AKIs, demonstrating the reasoning process). These issues are incorporated into a "prosocial" score (PScore, Eq. 1) to measure the complementary nature of the implementation. Note that the system must address at least one issue to have permission to run.

$$PScore = i_1\alpha_1 + \cdots + i_n\alpha_n \quad (1)$$

where $i_1$ is issue 1 (health professional shortages), $\alpha_1$ is the weight of $i_1$, $i_2$ is a secondary issue (unavailable reasoning), $\alpha_2$ is the weight of $i_2$, etc. Two options are available for each issue (true or false), where the value (ranges from 0-1) for true is 0.99 and false is 0.01. Note that the higher the value the more "prosocial" the option. Further, $\alpha_1$, ..., $\alpha_n$, must sum to 1. In this work, all issues are equally weighted (i.e., 0.333). Thus, the minimum PScore required for the system to run is 0.336. As we identified and addressed three issues (options set to true), the PScore was 0.989. This approach can be used to address a single or multiple issues.

**Pre-trained LLMs, fine-tuning, and retrieval-augmented generation.** LLMs in STRUC-MAS exploit the global structure (Appendix 6.1, Table 1) to infer local structures (i.e., individualized diagnostic reasoning) and diagnoses over time (i.e., iterations of iteration). We designed two paths: 1) *structure-following*, which exploits the global structure; and 2) *semi-structure-following*, which exploits the global structure and explores via retrieval-augmented generation (RAG) [18] – enabling agents to retrieve a previous similar note from the training set for comparison. Three open-source LLMs ranging from 8-32B parameters were selected based on their reported leading instruction-following and reasoning capabilities: 1) QWEN 2.5 Instruct [19]; Phi4 [20]; and Llama 3.1 Instruct [21], and fine-tuned via quantized low-rank adaptation (QLoRA) [22].

**Smart rounds.** To efficiently orchestrate multiple LLMs we build on existing frameworks and design a coordination style that is more akin to health system routines (see Appendix 6.2, Fig. 4) [23]. This setup mimics clinical rounds, where there is frequently a mixture of expertise levels (e.g., medical students, residents, attendings), to teach less experienced experts to perform at the level of more experienced experts (i.e., a mixture of experts) [24]. In this scenario, the assumption is that individual weaker experts can adopt the knowledge from stronger/more experienced experts to update their beliefs over time/rounds (i.e., knowledge distillation) [25]. Further, stronger experts can reinforce and/or increase their confidence in their beliefs. Likewise, smart rounds aim to optimize reaching consensus across the agents in as few rounds as possible without sacrificing performance. If performance gain in a given round is less than a set threshold then early stopping occurs (i.e., the agents achieved near maximum exploitation of the global structure).

**Multi-agent system logs/records and evaluation.** Multiagent records (MAR) log agent-generated output using a new vocabulary called agent-based terms (ABT) [26]. (see Appendix 6.3, Table 2-3). We evaluated the agent-based diagnosis (AD) using confusion matrix-based metrics (area under receiver operating characteristic curve, AUCROC; average precision, AP; precision; recall; false positive, FP; false negative, FN; true positive, TP; true negative, TN). We evaluated the agent-based diagnostic reasoning (ADR) using semantic similarity metrics (BERTScore) [27]. We developed a balanced classification and reasoning score (BCRScore, Eq. 2) to jointly assess AD and ADR (see Appendix 6.8, Table 8):

$$BCRScore = A\alpha + B\beta \quad (2)$$

where A is the AD metric (e.g., AP), $\alpha$ is A's weight, B is the ADR metric (e.g., average BERTScore F1), and $\beta$ is B's weight. Note that $\alpha$ and $\beta$ must sum to 1. Further, we assess agent confidence levels (ACL) as well as agent documentation burden via agent time spent on documentation (ATSD) and agent documentation length (ADL).

## 3 Results

Round 0 shows SF-FT Agents 2 and 3 perform similarly with AP of 0.186 and 0.202, respectively (see Appendix 6.4, Table 4). Notably, in Round 0, SF-FT Agent 1 performed poorly (AP of 0.133) with a recall of 0.01 (misdiagnosing almost all TP as FN), whereas SF-FT Agents 2 and 3 achieved a recall of 0.67 and 0.62, respectively. Round 1 shows SF-FT Agents 1, 2, and 3 perform similarly (AP of 0.190, 0.192, 0.198, respectively). Interestingly, via explicit interactions, SF-FT Agents 1 and 2 recall increased to 0.62 and 0.68, respectively (i.e., knowledge distillation). Similarly, SF-FT-RAG Agent 1 and Agent 2 recall increased to 0.52 and 0.69, respectively. Further, the recall for the BPRV increased from 0.37 to 0.62 in Round 1 for SF-FT (the team captures approximately 1.6 TP for every FN). Note that since P (round 1 BPRV AP) – O (round 0 BPRV AP) < Q (0.040), early stopping occurred.

Markedly, SF-FT Agent 1 was highly confident in Round 0 regarding TP and FN cases even though it was misdiagnosing almost all positive cases as negative cases, but after explicit interactions, its high confidence in TP and FN cases dropped (see Appendix 6.6, Table 6). We also observed that Agents 2 and 3 increased their confidence in TP and FN cases from Round 0 to Round 1, suggesting the explicit interactions reinforced their prior beliefs (see Appendix 6.6, Table 6). Appendix Table 7 shows that incorporating RAG made all agents highly confident in Round 0 TP and FN cases. Appendix Figure 5-6 shows the ADR alignment analysis by TP, FP, FN, and TN cases, highlighting that SF-FT Agent 3 reference group SF-FT Agent 2 were more aligned than SF-FT Agent 1 reference group SF-FT 2 in Round 0, while in Round 1 the alignment increased. Appendix Figure 7-8 shows that the SF-FT-RAG Agent 2 ADR was more aligned with SF-FT Agent 2 for TP, FP, FN, and TN cases. Appendix Table 5 shows the agent documentation burden. Appendix Table 8 shows the BCRScores.

## 4   Discussion

In this work, we highlight the role global structure plays in LLM-based agents. In our use case for predicting AKI, pretraining, fine-tuning, and/or RAG are no replacement for learning structure via traditional machine learning or statistics methods. In reviewing the "reasoning" for structure-following agents, we found that it leveraged the global structure of the data and associations between variables to provide local answers. Board configurations may be suitable for other clinical specialties as well, such as where the overlap of organs/systems makes it difficult to assume or know the underlying structures (e.g., oncology, cardiology, endocrinology, gastroenterology, rheumatology). Structure-following approaches seem like a promising path forward to utilize and evaluate these emerging technologies in the health domain. Future work may explore implementing multiple boards (i.e., multi-stakeholders) and dynamic resource allocation.

## 5   Conclusion

We demonstrated that global structure is necessary prior to achieving competitive classification and reasoning performance in the health domain. Further, we showed that not all models can leverage the global structure in a meaningful way; however, models that are stronger in specific capabilities can help the team improve. We showed that across multiple rounds agents can increase and decrease their confidence levels based on explicit interactions with other models. We also demonstrated that smart rounds were sufficient to get cheaper, faster, and informative results.

**Author Contributions.** Most intellectual work was done by DG with guidance from AB. DG planned and implemented the paper methodologies. DG created the software and multiagent records as well as ran the experiments and evaluated the models. PP, SN, and AB reviewed drafts of the paper and provided feedback.

# 6 Appendix

## 6.1 Learning the Structure Template

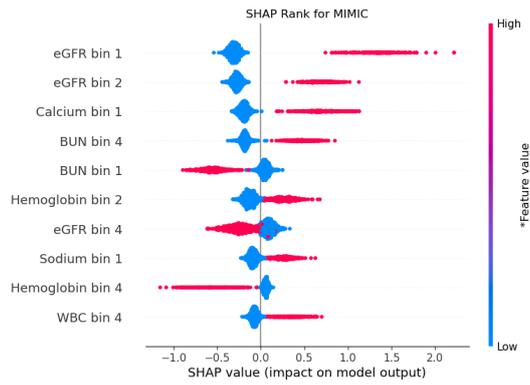

**Fig. 2.** SHAP value feature rank for MIMIC.

**Table 1.** Autogenerated Structure Template

| Site | Structure Template |
|---|---|
| MIMIC | Having the lowest bin (i.e., 1) for estimated glomerular filtration rate (eGFR) is the most important feature and indicates the highest risk for acute kidney injury (AKI). Having the second lowest bin (i.e., 2) for eGFR is the second most important feature and indicates higher risk for AKI. Having the lowest bin for calcium (i.e., 1) is the third most important feature and indicates higher risk for AKI. Having the highest bin for blood urea nitrogen (i.e., 4) is the fourth most important feature and indicates higher risk for AKI. Having the lowest bin for blood urea nitrogen (i.e., 1) is the fifth most important feature and indicates decreased risk for AKI. Having the second lowest bin for hemoglobin (i.e., 2) is the sixth most important feature and indicates higher risk for AKI. Having the highest bin for eGFR (i.e., 4) is the seventh most important feature and indicates decreased risk for AKI. Having the lowest bin for sodium (i.e., 1) is the eighth most important feature and indicates higher risk for AKI. Having the highest bin for hemoglobin (i.e., 4) is the ninth most important feature and indicates decreased risk for AKI. Having the highest bin for white blood cell count (i.e., 4) is the tenth most important feature and indicates increased risk for AKI. |

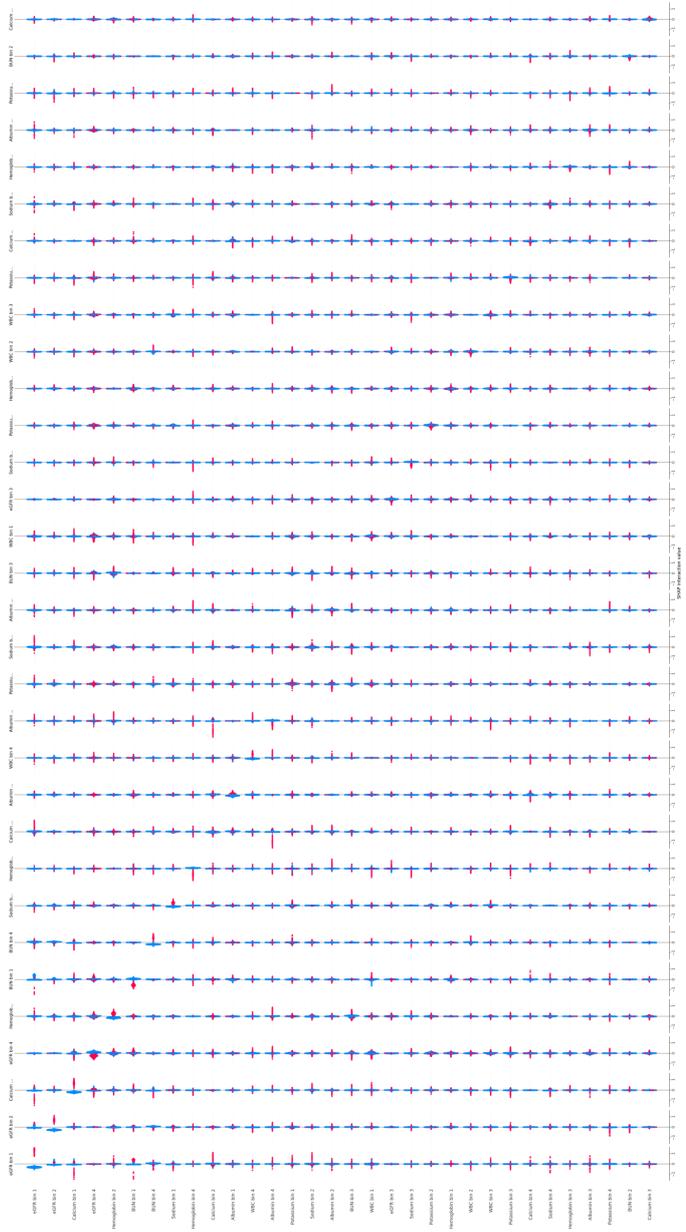

**Fig. 3.** SHAP interaction values for MIMIC. Main effects are on the diagonal and interaction effects are off-diagonal.

## 6.2 Smart Rounds

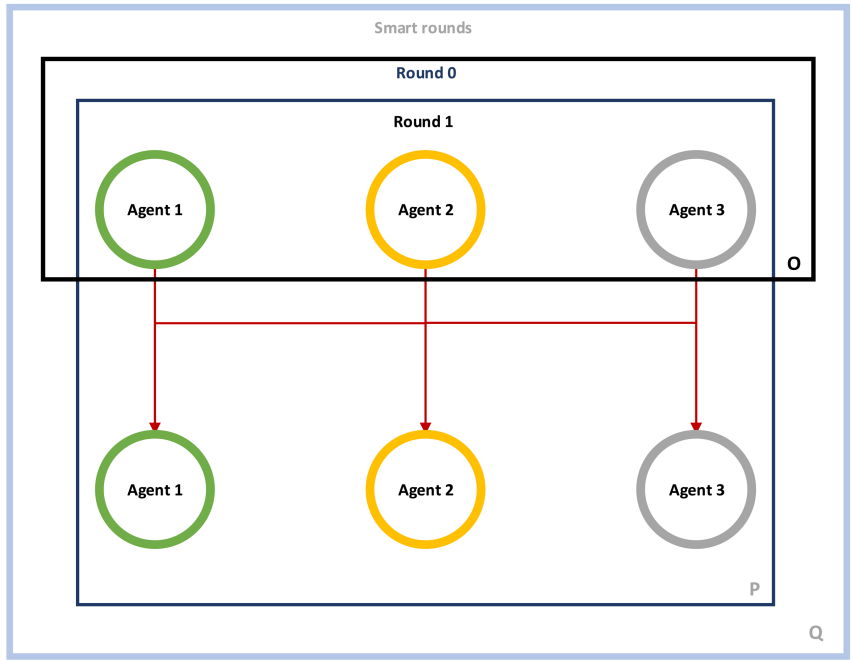

**Fig. 4.** Smart rounds: Include an outer plate (Round 0) and an inner plate (Round 1). Note that the inner plate repeats until P-O < Q. Q is the smart rounds early stopping threshold (e.g., Q = 0.040), P is the round 1 AP, and O is round 0 AP. In round 0 (implicit interactions), experts (i.e., agents) do their own individual assessments, and the end user (e.g., healthcare provider) may then calculate the aggregate/population statistics across the multiple agents via majority vote, precision-weighted vote, recall-weighted vote, or balanced precision-weighted-recall-weighted vote (BPRV). In round 1+ (explicit interactions), agents explicitly interact/consult with each other via conversational interactions (directed edges) to potentially update their prior diagnosis and prior diagnostic reasoning from round 0. This results in updated diagnosis, diagnostic reasoning and aggregate statistics.

## 6.3 Multi-Agent Records Agent-Based Terms (MAR-ABT)

**Table 2.** Multiagent records agent-based terms. Consists of standard terms, such as Agent Name (AN), Agent Diagnosis (AD), Agent Diagnostic Reasoning (ADR), Agent Confidence Level (ACL), Agent Documentation Length (ADL), and Agent Time Spent on Documentation (ATSD).

| | Agent Name (AN) | Agent Diagnosis (AD) | Agent Diagnostic Reasoning (ADR) | Agent Confidence Level (ACL) | Agent Documentation Length (ADL) | Agent Time Spent on Documentation (ATSD) |
|---|---|---|---|---|---|---|
| Reference | Structure-Following Agent 1 | 1.01, AKI | The patient has several indicators that suggest a higher risk for acute kidney injury (aki). the egfr is in the second lowest bin (2), indicating a higher risk. additionally, hemoglobin is in the second lowest bin (2) and sodium is in the lowest bin (1), both of which also indicate higher risk. while some factors like white blood cell count being in the third bin do not strongly suggest aki, the combination of multiple high-risk indicators suggests that there's a moderate likelihood of developing aki within two days. | High | 1521 tokens | 11 Seconds |
| Reference | Structure-Following Agent 2 | 1.01, AKI | The patient has multiple indicators (egfr at 2, hemoglobin at 2, sodium at 1, and calcium at 1) that are associated with a higher risk for acute kidney injury. these factors collectively suggest an increased likelihood of developing aki within two days. | High | 1872 tokens | 43 Seconds |
| Reference | Structure-Following Agent 3 | 1.01, AKI | The patient has several key indicators that suggest a higher risk for acute kidney injury (aki). the hemoglobin is in the second lowest bin (2), indicating a higher risk. additionally, sodium and calcium are both in their respective lowest bins (1), which also indicate higher risk. these factors collectively increase the likelihood of developing aki within two days. | High | 1486 tokens | 44 Seconds |

**Table 3.** Mapping (source) agent-based terms (ABT) vocabulary to (target) other standard vocabularies

| Vocabulary Mapping | Source | Target |
|---|---|---|
| AN → PN | Structure_Following_Agent_2 | Smith, Jane MD |
| AD → ICD-9 | AD: 1.01 | ICD-9: 589.4 |
| AD → ICD-10 | AD: 1.01 | ICD-10: N17.9 |
| AD → SNOMED | AD: 1.01 | SNOMED: 14003100011903 |
| ADR → Clinical Note (CN) | The patient has multiple indicators (egfr at 2, hemoglobin at 2, sodium at 1, and calcium at 1) that are associated with a higher risk for acute kidney injury; these factors collectively suggest an increased likelihood of developing aki within two days. | The patient's chief complaint was weakness, labs show low egfr, sodium, and calcium. Monitor labs for changes. Potential electrolyte imbalance, consult nephrology. |

## 6.4 Agent Diagnosis Performance Evaluations

Table 4. Structure-following agents vs. Non-structure-following agents (baselines)

| Round | Models | Structure-following fine-tuned (SF-FT) | | | | Structure-following fine-tuned retrieval-augmented generation (SF-FT-RAG) | | | | Non-structure following fine-tuned (NSF-FT baseline) | | | | Non-structure following fine-tuned retrieval-augmented generation (NSF-FT-RAG baseline) | | | |
|---|---|---|---|---|---|---|---|---|---|---|---|---|---|---|---|---|---|
| | | AUC[h] | AP | Pr.[d] | Re.[e] | AUC[h] | AP | Pr.[d] | Re.[e] | AUC[h] | AP | Pr.[d] | Re.[e] | AUC[h] | AP | Pr.[d] | Re.[e] |
| 0 | Agent 1[a] | 0.504 | 0.133 | 0.30 | 0.01 | 0.552 | 0.153 | 0.242 | 0.20 | 0.50 | 0.131 | nan | 0.00 | 0.500 | 0.131 | nan | 0.0 |
| | Agent 2[b] | 0.647 | 0.186 | 0.21 | 0.67 | 0.673 | 0.201 | 0.232 | 0.69 | 0.55 | 0.144 | 0.150 | 0.72 | 0.618 | 0.173 | 0.205 | 0.57 |
| | Agent 3[c] | 0.666 | 0.202 | 0.25 | 0.62 | 0.667 | 0.213 | 0.282 | 0.54 | 0.50 | 0.131 | nan | 0.00 | 0.619 | 0.170 | 0.230 | 0.48 |
| | Majority Vote | 0.659 | 0.202 | 0.26 | 0.57 | 0.662 | 0.208 | 0.272 | 0.54 | 0.50 | 0.131 | nan | 0.00 | 0.632 | 0.200 | 0.30 | 0.41 |
| | Precision-Weighted Vote[f] | 0.505 | 0.135 | 0.43 | 0.01 | 0.556 | 0.167 | 0.371 | 0.15 | 0.50 | 0.131 | nan | 0.00 | 0.50 | 0.1312 | nan | 0.00 |
| | Recall-Weighted Vote[g] | 0.654 | 0.187 | 0.21 | 0.72 | 0.673 | 0.201 | 0.232 | 0.69 | 0.55 | 0.144 | 0.150 | 0.72 | 0.605 | 0.165 | 0.183 | 0.64 |
| | Balanced Precision-Weighted-Recall-Weighted Vote[i] | **0.580** | **0.161** | **0.32** | **0.37** | **0.615** | **0.184** | **0.302** | **0.42** | **0.53** | **0.138** | **nan** | **0.36** | **0.553** | **0.148** | **nan** | **0.32** |
| 1 | Agent 1[a] | 0.649 | 0.190 | 0.23 | 0.62 | 0.625 | 0.180 | 0.225 | 0.52 | 0.53 | 0.139 | 0.144 | 0.60 | 0.586 | 0.169 | 0.255 | 0.31 |
| | Agent 2[b] | 0.659 | 0.192 | 0.22 | 0.68 | 0.672 | 0.201 | 0.232 | 0.69 | 0.54 | 0.143 | 0.152 | 0.57 | 0.631 | 0.181 | 0.217 | 0.58 |
| | Agent 3[c] | 0.652 | 0.198 | 0.25 | 0.55 | 0.665 | 0.208 | 0.269 | 0.56 | 0.50 | 0.131 | 0.000 | 0.00 | 0.633 | 0.201 | 0.303 | 0.41 |
| | Majority Vote | 0.648 | 0.190 | 0.23 | 0.61 | 0.662 | 0.200 | 0.246 | 0.60 | 0.52 | 0.137 | 0.143 | 0.49 | 0.637 | 0.195 | 0.268 | 0.47 |
| | Precision-Weighted Vote[f] | 0.652 | 0.198 | 0.25 | 0.55 | 0.628 | 0.189 | 0.260 | 0.45 | 0.52 | 0.137 | 0.143 | 0.49 | 0.583 | 0.178 | 0.328 | 0.24 |
| | Recall-Weighted Vote[g] | 0.661 | 0.192 | 0.22 | 0.69 | 0.671 | 0.198 | 0.225 | 0.71 | 0.55 | 0.145 | 0.151 | 0.69 | 0.632 | 0.181 | 0.217 | 0.58 |
| | Balanced Precision-Weighted-Recall-Weighted Vote[i] | **0.656** | **0.195** | **0.24** | **0.62** | **0.650** | **0.194** | **0.243** | **0.58** | **0.54** | **0.141** | **0.147** | **0.59** | **0.608** | **0.180** | **0.272** | **0.41** |

[a]Meta Llama 3.1 8B Parameters; [b]Microsoft Phi4 14B Parameters; [c]Alibaba Cloud Qwen 2.5 IT 32B Parameters;
[d]Precision = TP/(TP+FP); [e] Recall = TP/(TP+FN); [f]Threshold set at 0.25; [g]Threshold set at 0.75; [h]AUCROC; [i](Precision-weighted vote + Recall-weighted vote) /2.

## 6.5 Agent Documentation Burden

**Table 5.** SF-FT Agent Time Spent on Documentation (ATSD) and Agent Documentation Length (ADL)

| Round | Models | Agent Time Spent on Documentation (MM:SS.mmm or Min:Sec.Msec) | | | | | Agent Documentation Length (Total tokens) | | | | |
|---|---|---|---|---|---|---|---|---|---|---|---|
| | | Min. | 25% | 50% | 75% | Max. | Min. | 25% | 50% | 75% | Max. |
| 0 | Agent 1 | 0.0017 | 8.476 | 10.640 | 11.961 | 18.188 | 570 | 642 | 716 | 758 | 975 |
| | Agent 2 | 0.0025 | 28.593 | 30.680 | 32.481 | 49.646 | 783 | 907 | 937 | 966 | 1196 |
| | Agent 3 | 0.0004 | 34.302 | 39.579 | 1:13.494 | 2:40.391 | 578 | 617 | 630 | 645 | 731 |
| 1 | Agent 1 | 0.0024 | 9.864 | 11.327 | 13.531 | 23.612 | 1173 | 1360 | 1461 | 1567 | 2018 |
| | Agent 2 | 0.0018 | 37.221 | 40.428 | 43.616 | 58.986 | 1406 | 1697 | 1792 | 1883 | 2235 |
| | Agent 3 | 0.0010 | 46.841 | 53.270 | 1:23.257 | 3:07.440 | 1186 | 1366 | 1440 | 1521 | 1830 |

In Table 5 above we see the median agent time spent on documentation ranges from 11-40 seconds in round 0 to 11-53 seconds in round 1. Further, we see the median agent documentation length ranged from 630-937 tokens in round 0 to 1440-1792 tokens in round 1.

## 6.6 Agent Confidence Levels

**Table 6.** SF-FT Agents reported confidence levels by TP, FP, TN, and FN.

| Round | Model | Reported Confidence Level | TP (%) | FP (%) | TN (%) | FN (%) |
|---|---|---|---|---|---|---|
| 0 | Agent 1 | High (0.68 - 1) | 100% | 57% | 91% | 96% |
| | | Medium/Moderate (0.34 – 0.67) | 0% | 43% | 9% | 4% |
| | | Low (0 - 0.33) | 0% | 0% | 0% | 0% |
| | Agent 2 | High (0.68 - 1) | 64% | 42% | 59% | 46% |
| | | Medium/Moderate (0.34 – 0.67) | 36% | 58% | 41% | 53% |
| | | Low (0 - 0.33) | 0% | 0% | 0% | 1% |
| | Agent 3 | High (0.68 - 1) | 51% | 23% | 0% | 0% |
| | | Medium/Moderate (0.34 – 0.67) | 49% | 77% | 10% | 11% |
| | | Low (0 - 0.33) | 0% | 0% | 90% | 89% |
| 1 | Agent 1 | High (0.68 - 1) | 86% | 80% | 87% | 88% |
| | | Medium/Moderate (0.34 – 0.67) | 14% | 20% | 10% | 11% |
| | | Low (0 - 0.33) | 0% | 0% | 3% | 1% |
| | Agent 2 | High (0.68 - 1) | 97% | 91% | 94% | 89% |
| | | Medium/Moderate (0.34 – 0.67) | 3% | 9% | 6% | 11% |
| | | Low (0 - 0.33) | 0% | 0% | 0% | 0% |
| | Agent 3 | High (0.68 - 1) | 94% | 91% | 63% | 58% |
| | | Medium/Moderate (0.34 – 0.67) | 6% | 9% | 37% | 42% |
| | | Low (0 - 0.33) | 0% | 0% | 0% | 0% |

Table 6 above shows that in the initial round Agent 1 was highly confident in its diagnosis and ADR of FN cases; however, Agent 1 missed almost all the positive cases, suggesting it couldn't differentiate between the groups very effectively. After explicit interactions in round 1 we see that Agent 1 reduced its high confidence, suggesting the other agent's diagnosis and ADR was influential. Further, we see that in the initial round Agent 2 and Agent 3, though having considerably higher recall than Agent 1, were more uncertain and after explicit interactions with the other agent's their confidence levels increased.

**Table 7.** SF-FT-RAG and NSF-FT-RAG Agents reported confidence levels by TP and FN.

| Round | Model | Reported Confidence Level | TP | | FN | |
|---|---|---|---|---|---|---|
| | | | SF-FT-RAG (%) | NSF-FT-RAG (%) | SF-FT-RAG (%) | NSF-FT-RAG (%) |
| 0 | Agent 1 | High (0.68 - 1) | 100% | nan | 100% | 100% |
| | | Medium/Moderate (0.34 – 0.67) | 0% | nan | 0% | 0% |
| | | Low (0 - 0.33) | 0% | nan | 0% | 0% |
| | Agent 2 | High (0.68 - 1) | 100% | 100% | 100% | 100% |
| | | Medium/Moderate (0.34 – 0.67) | 0% | 0% | 0% | 0% |
| | | Low (0 - 0.33) | 0% | 0% | 0% | 0% |
| | Agent 3 | High (0.68 - 1) | 100% | 100% | 100% | 100% |
| | | Medium/Moderate (0.34 – 0.67) | 0% | 0% | 0% | 0% |
| | | Low (0 - 0.33) | 0% | 0% | 0% | 0% |
| 1 | Agent 1 | High (0.68 - 1) | 96% | 77% | 28% | 2% |
| | | Medium/Moderate (0.34 – 0.67) | 4% | 23% | 72% | 98% |
| | | Low (0 - 0.33) | 0% | 0% | 0% | 0% |
| | Agent 2 | High (0.68 - 1) | 100% | 100% | 100% | 100% |
| | | Medium/Moderate (0.34 – 0.67) | 0% | 0% | 0% | 0% |
| | | Low (0 - 0.33) | 0% | 0% | 0% | 0% |
| | Agent 3 | High (0.68 - 1) | 100% | 100% | 100% | 99% |
| | | Medium/Moderate (0.34 – 0.67) | 0% | 0% | 0% | 1% |
| | | Low (0 - 0.33) | 0% | 0% | 0% | 0% |

## 6.7 Agent Diagnostic Reasoning Alignment Analysis

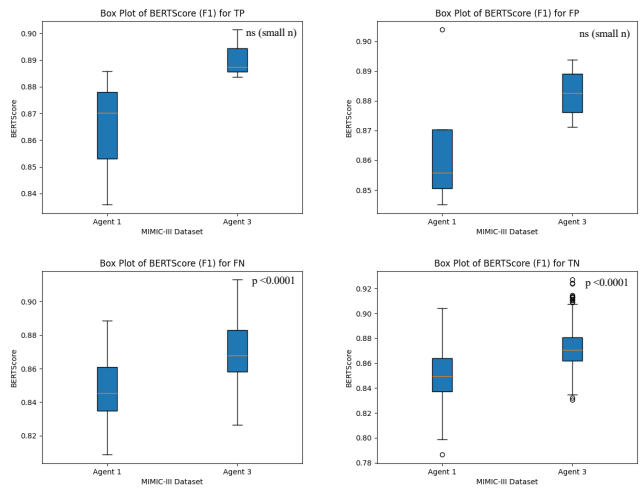

**Fig. 5.** SF-FT Round 0

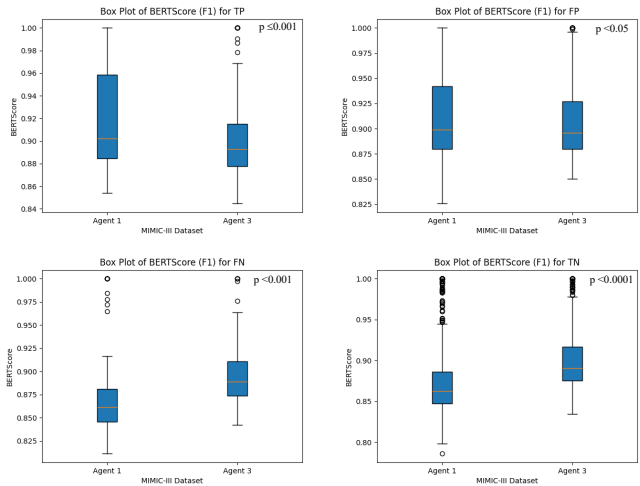

**Fig. 6.** SF-FT Round 1

**Fig. 5-6.** In Fig. 5 above we see Agent 3 reference group Agent 2 (highest recall) is more aligned than Agent 1 reference group Agent 2. Since Agent 2 and Agent 3 have similar precision and recall scores we expect those two agents to be more aligned across case types. In Fig. 6 we see that after explicit interactions the BERTScore increases. Further, we see that the gap between Agent 3 reference group Agent 2 and Agent 1 reference group Agent 2 decreases for TP and FP cases, suggesting more similar ADR. Yet, they are still statistically significant different, suggesting their ADR are not identical (i.e., there exists some unique reasoning).

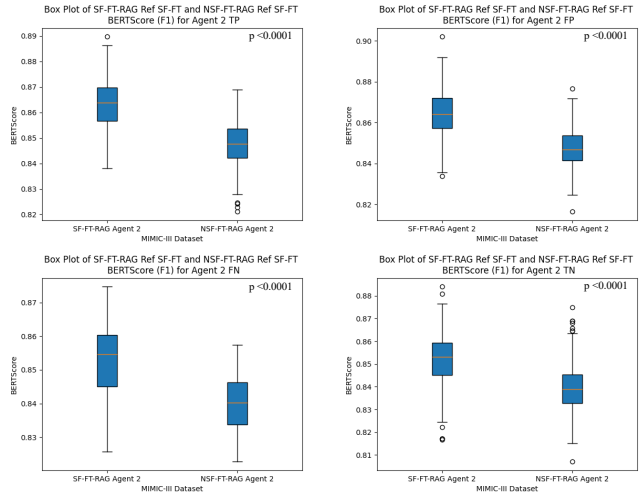

**Fig. 7.** SF-FT-RAG Agent 2 reference group SF-FT Agent 2 and NSF-FT-RAG Agent 2 reference group SF-FT Agent 2 Round 0

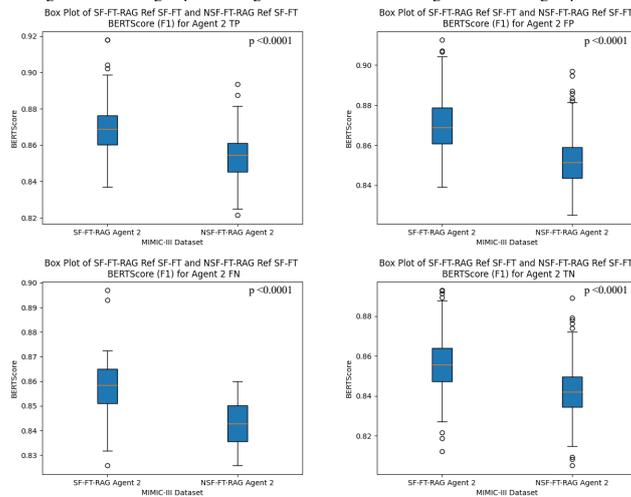

**Fig. 8.** SF-FT-RAG Agent 2 reference group SF-FT Agent 2 and NSF-FT-RAG Agent 2 reference group SF-FT Agent 2 Round 1

**Fig. 7-8.** In Fig. 7 above we see SF-FT-RAG Agent 2 reference group SF-FT Agent 2 (highest SF-FT recall agent) is more aligned than NSF-FT-RAG Agent 2 reference group SF-FT Agent 2. We hope to see this trend continue in subsequent rounds to demonstrate the utility of structure-following combined with RAG (i.e., semi-structure-following) vs. RAG alone. In Fig. 8 we see that after explicit interactions the BERTScores increase and SF-FT-RAG Agent 2 reference group SF-FT Agent 2 remains more aligned than NSF-FT-RAG Agent 2 reference group SF-FT Agent 2, demonstrating the benefit of leveraging the global structure.

## 6.8 Joint Assessment of Classification and Reasoning

Table 8. Balanced Classification and Reasoning Score (BCRScore)

| AD Metric | α | ADR Metric | β | Model | Case Type | BCRScore Round 0 | BCRScore Round 1 |
|---|---|---|---|---|---|---|---|
| AP | 0.5 | Average BERTScore F1 | 0.5 | SF-FT Agent 1 Ref. SF-FT Agent 2 | FN | (0.161)*0.5+ (0.846)*0.5 =0.5035 | (0.195)*0.5+ (0.878)*0.5 =0.5365 |
| AP | 0.5 | Average BERTScore F1 | 0.5 | SF-FT Agent 1 Ref. SF-FT Agent 2 | TP | (0.161)*0.5+ (0.864)*0.5 =0.5125 | (0.195)*0.5+ (0.919)*0.5 =0.5570 |
| AP | 0.5 | Average BERTScore F1 | 0.5 | SF-FT-RAG Agent 2 Ref. SF-FT Agent 2 | FN | (0.184)*0.5+ (0.853)*0.5 = 0.5185 | (0.194)*0.5+ (0.858)*0.5 =0.5260 |
| AP | 0.5 | Average BERTScore F1 | 0.5 | NSF-FT-RAG Agent 2 Ref. SF-FT Agent 2 | FN | (0.148)*0.5+ (0.841)*0.5 =0.4945 | (0.180)*0.5+ (0.843)*0.5 =.5115 |

The BCRScore aims to provide a balance between the agent classification performance and the agent diagnostic reasoning. In this implementation, we set α and β to 0.50. However, end-users may tune α and β to favor classification over reasoning and vice versa, as long as they sum to 1. Note that while we use the average precision (AP) of the balanced precision-weighted-recall-weighted vote for the AD metric, it can be replaced with other metrics (e.g., AUCROC, etc.). Also, regarding the ADR metric, we selected the average BERTScore F1, but it can be replaced with other metrics (e.g., (average) Rouge, (average) BLEU, etc.). Note that the higher the BCRScore the better the overall performance for diagnosis as well as diagnostic reasoning alignment. In Table 8 above we see that the BCRScore in round 1 was higher than in round 0 for SF-FT Agent 1 reference group SF-FT Agent 2 for FN cases, suggesting improvement in agent diagnosis and agent diagnostic reasoning for FN cases. Similarly, we see that the BCRScore in round 1 was higher than in round 0 for SF-FT Agent 1 reference group SF-FT Agent 2 for TP cases.